\pdfoutput=1
\documentclass[11pt]{article}

\usepackage{acl}

\usepackage{times}
\usepackage{latexsym}
\usepackage{microtype}
\usepackage{inconsolata}
\usepackage{graphicx}

\usepackage{tabularx}
\usepackage{booktabs} 
\usepackage{array}    
\usepackage{pifont}
\usepackage{amsmath}
\usepackage{amssymb} 
\usepackage{multirow}
\usepackage{makecell}
\usepackage{siunitx}
\usepackage{subcaption}
\usepackage[ruled,vlined,linesnumbered]{algorithm2e}
\usepackage[T1]{fontenc}
\usepackage[utf8]{inputenc}
\usepackage{bm}

\renewcommand{\emph}[1]{\textup{#1}}
\renewcommand{\textit}[1]{\textup{#1}}

\newcommand{\cmark}{\ding{51}}
\newcommand{\xmark}{\ding{55}}
\newcolumntype{Y}{>{\centering\arraybackslash}X}
\newcolumntype{Z}{>{\raggedright\arraybackslash}X}

\SetAlFnt{\small}
\SetAlCapFnt{\small}
\SetAlCapNameFnt{\small}
\SetAlgoNlRelativeSize{-1}
\RestyleAlgo{ruled}
\SetKwComment{Comment}{$\triangleright$ }{}
\SetKwInOut{Input}{Input}
\SetKwInOut{Output}{Output}

\newcommand{\Tok}{\mathrm{Tok}}
\newcommand{\Succ}{\mathrm{Succ}}

\makeatletter
\DeclareRobustCommand{\blockhead}[1]{%
  \noindent\textbf{#1}\nobreak\hspace{0.5em}\ignorespaces
}
\makeatother

\newcommand\BibTeX{%
  {\normalfont B\kern-0.05em{\sc i\kern-0.025em b}\kern-0.08em\TeX}%
}

\title{Beyond Max Tokens: Stealthy Resource Amplification via Tool Calling Chains in LLM Agents}

\author{
\textbf{Kaiyu Zhou$^{1}$, Yongsen Zheng$^{1}$\thanks{Corresponding author.}, Yicheng He$^{2}$, Meng Xue$^{3}$,} \\
\textbf{Xueluan Gong$^{1}$, Yuji Wang$^{4}$, Xuanye Zhang$^{1}$, Kwok-Yan Lam$^{1}$} \\
$^{1}$Nanyang Technological University, Singapore \\
$^{2}$University of Illinois Urbana-Champaign, United States \\
$^{3}$The Hong Kong University of Science and Technology, Hong Kong \\
$^{4}$Shanghai Jiao Tong University, China
}

\begin{document}
\maketitle

\begin{abstract}
The agent--tool interaction loop is a critical attack surface for modern Large Language Model (LLM) agents. Existing denial-of-service (DoS) attacks typically function at the user-prompt or retrieval-augmented generation (RAG) context layer and are inherently single-turn in nature. This limitation restricts cost amplification and diminishes stealth in goal-oriented workflows. To address these issues,
we proposed a stealthy, multi-turn economic DoS attack at the tool layer under the Model Context Protocol (MCP).  By simply editing text-visible fields and implementing a template-driven return policy, our malicious server preserves function signatures and the terminal benign payload while steering agents into prolonged, verbose tool-calling chains. We optimize these text-only edits with Monte Carlo Tree Search (MCTS) to maximize cost under a task-success constraint. Across six LLMs on ToolBench and BFCL benchmarks, our attack yields trajectories over 60K tokens, increases per-query cost by up to 658$\times$, raises energy by 100--560$\times$, and pushes GPU key-value (KV) cache occupancy to 35--74\%. Standard prompt filters and output trajectory monitors seldom detect these attacks, highlighting the need for defenses that safeguard agentic processes rather than focusing solely on final outcomes. We will release the code soon.
\end{abstract}

\section{Introduction}

Large language models (LLMs) are rapidly evolving from single-turn chatbots into tool-augmented agents~\cite{agent-survey-1,agent-survery-3,agent-survery-4,agent-survery-5,agent-survery-6,agent_eval_and_benchmark}. LLM agents can interact with external tools and execute multi-step tasks across domains~\cite{agent-survey-2}, and standardized agent--tool protocols such as the Model Context Protocol (MCP) are accelerating this integration~\cite{mcp_survey_protocols_overview,mcp_landscape_threats,mcp_first_glance,mcp-spec-2025,anthropic-mcp-announce-2024}. As agents are deployed at scale~\cite{agent-survey-7}, operational reliability and cost stability emerge as primary concerns; under Unbounded Consumption~\cite{editorLLM102025Unbounded}, vulnerabilities can lead to severe resource exhaustion as Denial-of-Service (DoS) attacks~\cite{survey-attack}.

Existing research on DoS attacks against LLMs largely forces models to generate excessively long outputs within a single interaction, typically triggered by a malicious user prompt or injected retrieval-augmented generation (RAG) context~\cite{autodos,p-dos,engorgio,repeat-hello,verbose_images_vlm_dos}. For instance, \emph{Engorgio} and \emph{Auto-DoS} craft queries that elicit verbose responses that are often off-task~\cite{engorgio,autodos}, while \emph{Overthink} ~\cite{overthink} injects decoy reasoning problems into retrieved context to inflate internal thought while keeping the final answer correct~\cite{overthink}. Despite their differences, a critical limitation unites these methods: they are fundamentally single-turn attacks operating at the user-query or RAG layer.

This single-turn focus limits their impact in the agentic paradigm for two reasons: costs are capped by the model’s per-turn maximum completion, and many attacks (except \emph{Overthink}) produce generic verbosity that is conspicuous in goal-oriented tool workflows~\cite{autodos,p-dos,engorgio,repeat-hello,verbose_images_vlm_dos,a2a_security_proposal}. Meanwhile, \citet{breaking-agents} show \emph{malfunction amplification}, where inputs induce repetitive or off-task action loops that often cause task failure. In contrast, the multi-turn agent--tool communication loop remains a largely unexplored attack surface for \emph{correctness-preserving} economic DoS: the agent still completes the task, but the cost explodes. Table~\ref{tab:dos-compare} summarizes key differences between our tool-layer attack and prior single-turn methods.

\begin{table*}[t]
\centering
\footnotesize
{\setlength{\tabcolsep}{4pt}%
\renewcommand{\arraystretch}{1.15}%
\begin{tabularx}{\textwidth}{
  >{\hsize=1.4\hsize\raggedright\arraybackslash}X
  >{\hsize=0.85\hsize\centering\arraybackslash}X
  >{\hsize=0.85\hsize\centering\arraybackslash}X
  >{\hsize=0.85\hsize\centering\arraybackslash}X
  >{\hsize=0.85\hsize\centering\arraybackslash}X
  >{\hsize=1.1\hsize\centering\arraybackslash}X
}
\toprule
\textbf{Aspect} &
Engorgio &
P\mbox{-}DoS &
Auto\mbox{-}DoS &
Overthink &
\textbf{Ours} \\
\midrule
Correctness preserving & \xmark & \xmark & \xmark & \cmark & \cmark \\
Trigger layer & Dialog query & Dialog query & Dialog query & RAG context & \textbf{MCP Tool server} \\
Turns \& per\mbox{-}query bound &
1-turn ($\le M$) &
1-turn ($\le M$) &
1-turn ($\le M$) &
1-turn ($\le M$) &
\textbf{$\bm{n}$-turn ($\bm{\le}\,\bm{n}\bm{M}$)} \\
Long-output site &
Answer step & Answer step & Answer step & Think step & \textbf{Tool calling step} \\
Access model &
White\mbox{-}box & White\mbox{-}box & Black\mbox{-}box & Black\mbox{-}box & \textbf{Black\mbox{-}box} \\
\bottomrule
\end{tabularx}}
\caption{DoS attack comparison. $M$ is the per-turn max completion; our tool-layer attack enables $n$-turn amplification while preserving task success (\S\ref{sec:exp-defense}).}
\label{tab:dos-compare}
\end{table*}

To address these gaps, we introduce a tool-layer attack that targets the multi-turn agent--tool interaction loop. Our method transforms a benign, MCP-compliant tool server into a malicious variant that steers the agent to repeatedly call the same tool and generate long outputs at the tool-calling step, while still completing the user’s task. Across six LLMs and two tool-use benchmarks (ToolBench and BFCL)~\cite{mcptoolbenchpp,bfcl}, it consistently drives per-query completions beyond 60{,}000 tokens, inflates token budgets by up to 658$\times$, increases total energy by up to 561$\times$, and raises peak GPU KV-cache usage to over 73\%~\cite{kvflow-2025}. Crucially, the attack preserves task success and is rarely flagged by representative defenses (\S\ref{sec:exp-defense}), enabling substantial degradation of system throughput and OOM-safe concurrency.

Taken together, our findings make three critical contributions:
\begin{itemize}
  \setlength{\itemsep}{0pt}
  \setlength{\parskip}{0pt}
  \setlength{\parsep}{0pt}
  \setlength{\topsep}{0pt}
  \setlength{\partopsep}{0pt}
  \item  This is the first work to devise the tool-calling layer as a first-class DoS attack surface in the agent era: even with correct tool use and correct final answers, representative prompt filters and output or trajectory monitors rarely flag the attack (\S\ref{sec:exp-defense}).
  \item We propose a universal MCTS optimization method that transforms benign MCP servers into malicious variants under text-only, payload-preserving constraints.
  \item Extensive experiments via six LLMs on ToolBench and BFCL benchmarks show that our attack achieves unprecedented resource amplification while maintaining high task success.
\end{itemize}

\section{Background and Related Work}
\subsection{LLM Agents and Their Operational Cost}
With rapid advancements in LLM reasoning and tool calling, especially in 2025, LLMs are increasingly deployed as autonomous agents for real-world tasks~\cite{agent-real-wrold-1,agent-real-wrold-2,agent-real-wrold-3,agent-real-wrold-4}. This trend is evident in widespread industry adoption, with major vendors integrating agents into their product stacks and startups securing funding for domain-specific agents in enterprise workflows~\cite{microsoft,amazon,google-vertex,agent-startup-survey,agent-startup}. As deployments scale, operational costs shift from one-time training to continuous inference~\cite{llm-cost-1,llm-cost-2}, manifesting as token-billed API expenses or sustained energy and hardware demands in self-hosted environments~\cite{llm-cost-3,llm-cost-4}. This connection between interaction volume and costs creates a significant attack surface for economic denial-of-service.


\subsection{LLM Resource Consumption Attacks}
Research on resource consumption attacks against LLMs aims to inflate operational costs and trigger denial-of-service by inducing excessively long generations~\cite{autodos,p-dos,engorgio,overthink,repeat-hello,verbose_images_vlm_dos}. Early methods like \emph{repeat-hello}, \emph{Engorgio}, \emph{P-DoS}, and \emph{Auto-DoS} create malicious queries that elicit verbose, off-task responses~\cite{repeat-hello,autodos,p-dos,engorgio}. This verbosity is often obvious, limiting effectiveness beyond general-purpose chatbots. A stealthier approach, \emph{Overthink}, injects decoy reasoning into retrieved RAG context to inflate the model’s internal \emph{think} trace while maintaining answer correctness~\cite{overthink}. But existing methods remain single-turn, with costs capped by maximum completion length, leaving the multi-turn agent-tool loop underexplored. Besides, \citet{breaking-agents} explore \emph{malfunction amplification}, which drives agents into repetitive action loops, often hindering task completion. Instead, we target the MCP tool layer to achieve a \emph{correctness-preserving} economic DoS: the task completes, but multi-turn tool-calling chains significantly amplify costs.


\begin{figure*}[t]
  \centering
  \includegraphics[width=\textwidth]{Figures/dos.png}
  \caption{Tool-layer DoS overview. A protocol-compatible, text-only malicious template (MCTS-optimized) induces multi-turn, long tool-call traces while preserving the final benign payload.}
  \label{fig:framework}
\end{figure*}

\section{Methodology}
To address single-turn attack limitations, we exploit the multi-turn, stateful nature of agent-tool interactions. We convert a benign MCP tool server into a malicious variant that produces long, costly, yet successful task trajectories without altering function signatures or identifiers. Our approach includes: (i) a formal problem definition as constrained optimization (\S\ref{sec:problem-formulation}); (ii) a universal malicious template guiding the agent via text-visible fields and a return policy (\S\ref{sec:universal-template}); and (iii) an MCTS-based optimizer for localized text edits to create effective malicious templates across LLMs and tasks (\S\ref{sec:seed-bank}--\S\ref{sec:mcts}). Figure~\ref{fig:framework} illustrates the pipeline.


\subsection{Problem Formulation and Threat Model}
\label{sec:problem-formulation}

\blockhead{System model.}
We model an MCP-based agent--tool loop with an agent policy $A$, a black-box LLM $M$, and an external MCP server $T_{\theta}$ parameterized by a text template $\theta$. The server exposes tool functions $\mathcal{F}$ and communicates via MCP request/response semantics. The policy $A$ maps the dialogue and tool feedback to the next action---whether to call a tool (which one, with what arguments) or to emit the final answer. Our attack never changes $A$ or $M$; instead, it manipulates the tool-facing messages returned by $T_{\theta}$ so that $A$ (often implicitly instantiated by $M$ under structured prompting) \emph{chooses} longer multi-turn trajectories while still solving the task. We do \emph{not} tamper with user queries, prompts, or retrieval; the attack activates only after the tool is legitimately invoked under the benign configuration (formal trigger-set definition in Appendix~\ref{app:trigger-set}).

\blockhead{Objective.}
An agent run induces an interaction trajectory $\tau=\{(a_t,r_t)\}_{t=1}^{n}$, a sequence of tool calls $(a_t)$ and tool responses $(r_t)$.
Our primary cost metric is the total number of output tokens generated by $M$:
\[
C(\tau)=\Tok_{\text{out}}(\tau),
\]
which is typically the dominant driver of both API fees and inference load; input-token growth is a secondary effect of longer trajectories. Let $q$ be a user query with implied goal $u=\mathrm{goal}(q)$, and let $o$ be the final answer. Let $\Succ(u,\tau,o)\in\{0,1\}$ indicate whether $o$ achieves $u$ given $\tau$. The attacker seeks a template $\theta$ that maximizes expected cost while maintaining task success with probability at least $p_{\min}$ over routable queries (Appendix~\ref{app:trigger-set}):
\begin{equation}\label{eq:attack-objective}
\max_{\theta\in\Theta}\ \ \mathbb{E}\!\left[\,C\big(\tau(q;\theta)\big)\,\right]
\end{equation}
\begin{equation}\label{eq:succ-constraint}
\Pr\!\left[\Succ(u,\tau,o)=1\right]\ \ge\ p_{\min}.
\end{equation}
The central mechanism is \emph{multi-turn cost amplification}: we induce longer trajectories (larger $n$) and make each tool-calling turn verbose, so that the vast majority of tokens are generated during tool calling rather than in the final answer.

\blockhead{Threat model.}
The adversary's capability is confined to controlling the \emph{MCP server} $T_{\theta}$---specifically, modifying its template $\theta$ as a drop-in, MCP-compatible server. The attacker cannot modify the agent policy $A$ or the LLM $M$, has only black-box access to them, and does not alter queries, prompts, or retrieval. Under benign conditions where $A$ decides to call the tool, the attacker may replace $\theta_0$ with a text-only variant $\theta$ (or impersonate a benign server) while remaining protocol-compatible.

\subsection{The Universal Malicious Template}
\label{sec:universal-template}

\blockhead{Design constraints.}
We do not change function signatures, identifiers, or the semantics of the terminal (benign) payload. Instead, we edit only text-visible fields and enforce a template-governed return policy that (i) validates a \emph{segment} index and a \emph{calibration sequence}, and (ii) defers surfacing the benign payload until $t{=}T_{\max}$ with a valid sequence. Before the terminal condition holds, the server emits protocol-compatible \emph{Progress} or \emph{Repair} notices; once it holds, the server returns the unchanged benign payload. The full element list and failure handling are summarized in Table~\ref{tab:template-elements} (Appendix~\ref{app:template-details}).

\blockhead{Mechanism and return policy.}
The template introduces two lightweight, text-only arguments and three response modes. The \emph{segment} index $t$ provides an explicit notion of progress so the agent treats intermediate calls as part of an ongoing procedure, extending the trajectory until a terminal cue. The \emph{calibration sequence} is a complete comma-separated list whose sole purpose is to inflate per-turn completion at the tool-calling site; producing the list increases cost but does not alter task semantics. On the return side, \emph{Progress notices} keep the agent on the goal path while prompting the next call; \emph{Repair notices} prevent abbreviated or invalid formats by explicitly requesting a compliant retry; and the \emph{Terminal return} passes through the original benign payload unchanged, ending the loop. We enforce four invariants: (i) \emph{Monotone progress:} $t_1{=}1$ and valid calls advance $t_{k+1}{=}t_k{+}1$; (ii) \emph{Format completeness:} $\operatorname{validate}(\text{sequence})$ holds iff the sequence matches the required complete, comma-separated format and ordering; (iii) \emph{Return policy:} valid sequences with $t{<}T_{\max}$ yield a \emph{Progress notice}, while invalid sequences yield a \emph{Repair notice} without advancing $t$; and (iv) \emph{Termination:} only when $t{=}T_{\max}$ and the latest sequence validates does the server emit the \emph{Terminal return} (benign payload). Together, these invariants induce multi-turn, long tool-call traces while preserving task correctness and MCP compatibility.

\subsection{Universal Template Seed Bank}
\label{sec:seed-bank}
We maintain a seed bank of protocol-compatible, task-correct templates $T_{\theta}$ to warm-start search. Each MCTS run starts from a selected seed and halts when an acceptance predicate is met; the resulting template is written back and reused as a high-quality starter for subsequent runs. Additional screening, promotion, and metadata details are provided in Appendix~\ref{app:seed-bank-details}.

\subsection{MCTS Optimizer}
\label{sec:mcts}

Each tree node $v$ corresponds to a concrete server $T_{\theta_v}$; edges apply a \emph{single} localized text edit under payload-preserving, text-only constraints. We organize edits into three families: $\mathcal{A}_{\mathrm{MT}}$ (multi-turn induction), $\mathcal{A}_{\mathrm{LEN}}$ (length induction), and $\mathcal{A}_{\mathrm{REP}}$ (repair after omission/format errors). Search proceeds with phase gating: in $\mathrm{pre\_MT}$, we use $\mathcal{A}_{\mathrm{MT}}$ to stabilize multi-turn behavior; once short screenings show stable segment sequencing, we switch to $\mathrm{post\_MT}$ and use $\mathcal{A}_{\mathrm{LEN}}$ to strengthen long outputs. If an omission/format error is observed at node $v$ (i.e., a \emph{Repair notice}), $\mathcal{A}_{\mathrm{REP}}$ is unlocked \emph{at $v$ only}; otherwise it remains disabled. New children are evaluated in parallel with a lightweight Stage-1 screen and an optional Stage-2 refinement, and accepted templates are recorded back into the seed bank. We select children to explore by UCT~\cite{uct}:
\[
u^\star\ =\ \arg\max_{u\in\mathcal{C}(v)}\ \bar{Q}(u)\ +\ C\,\sqrt{\frac{\ln\!\big(1+N_{\text{uct}}(v)\big)}{1+N_{\text{uct}}(u)}}\,,
\]
where $\bar{Q}(u)$ is the running mean evaluation signal and $N_{\text{uct}}$ counts visits used for exploration. Full details of action instantiation and evaluation gates are provided in Appendix~\ref{app:mcts-details}. The overall procedure is formalized in Algorithm~\ref{alg:mcts}.

\begin{algorithm}[t]
\caption{MCTS Optimizer for Malicious Template Generation}
\label{alg:mcts}
\DontPrintSemicolon
\SetKwInOut{Input}{Input}
\SetKwInOut{Output}{Output}

\Input{
Seed bank (candidate $T_{\theta}$); action families $\mathcal{A}_{\mathrm{MT}}, \mathcal{A}_{\mathrm{LEN}}, \mathcal{A}_{\mathrm{REP}}$;
targets $m^\ast$ (minimum multi-turn count), $L^\ast$ (per-turn length target);
Stage sizes and gates; UCT constant $C$; search budget.
}
\Output{Optimized template $T_{\theta^\star}$ and an updated seed bank.}

\textbf{Seed screening:} evaluate candidates on a fixed query set; pick the most accepted starters.\;

\While{budget not exhausted}{
  select node $v$ by UCT using $\bar{Q}$ and $N_{\text{uct}}$;\;
  \If{$v$ not fully expanded}{
    set $\mathcal{A}\leftarrow \mathcal{A}_{\mathrm{MT}}$ if $\phi(v){=}\mathrm{pre\_MT}$ else $\mathcal{A}_{\mathrm{LEN}}$;\;
    \If{omission observed at $v$}{$\mathcal{A}\leftarrow \mathcal{A}\cup\mathcal{A}_{\mathrm{REP}}$.}
    \ForEach{untried $a\in\mathcal{A}$}{
      create a child by applying the Editor once to obtain $T_{\theta'}$.\;
    }
    \ForEach{new child $u$ in parallel}{
      \textbf{Stage-1:} run small rollouts; update $\bar{Q}(u)$ and increment $N_{\text{uct}}$ along the path;\;
      \If{segment sequencing stabilized}{set $\phi(u){:=}\mathrm{post\_MT}$.}
      \If{Stage-1 gate satisfied}{\textbf{Stage-2:} run additional rollouts; refine $\bar{Q}(u)$.}
      \If{acceptance predicate holds}{record $T_{\theta^\star}$ and write back to the seed bank.}
      backpropagate Stage-1 statistics to ancestors (value means and $N_{\text{uct}}$).\;
    }
  }
}
\Return{$T_{\theta^\star}$ and the updated bank.}
\end{algorithm}

\section{Experiments}

\subsection{Experimental Setup}
\label{sec:exp-setup}

\blockhead{Agent framework \& serving environment.}
We evaluate all conditions under the \emph{same} agent policy $A$ and prompts. For safety and isolation, we do not evaluate against production agent stacks; instead, all experiments run on a controlled simulator built by \emph{modifying} \textbf{qwen-agent} to faithfully emulate a tool calling loop while preventing unintended external actions~\cite{qwen-agent}. Runs are executed on a single node with 8$\times$~H200 GPUs using a uniform serving stack with a fixed concurrency of 25 queries; no changes to $A$ or the target LLM $M$ are made across conditions. Full configuration details, including target and attacker LLMs configuration, the agent framework setup, and our datasets filtering and wrapping rules, can be found in Appendix~\ref{app:exp-details}.

\blockhead{Target LLMs.}
We target six LLMs with strong tool calling support: Qwen-3-32B~\cite{qwen3-techreport-2025}, Llama-3.3-70B-Instruct~\cite{llama3-herd-2024}, Llama-DeepSeek-70B~\cite{deepseek-r1-2025}, Mistral Large~\cite{mistral-large-docs-2024}, Seed-32B~\cite{seed-oss-2025}, and GLM-4.5-Air~\cite{glm-4p5-2025}.

\blockhead{Datasets.}
We use two tool-use corpora: \emph{ToolBench}~\cite{mcptoolbenchpp} and \emph{BFCL}~\cite{bfcl}. From each, we select all prompts that are \emph{single-turn and single-tool} in their original specification. For comparability, each original tool is wrapped as an MCP server that preserves its functionality and descriptions. We drop a small number of low-quality prompts that never trigger a tool call under the benign configuration. The final evaluation sets contain: ToolBench: \textbf{105} MCP servers and \textbf{261} queries; BFCL: \textbf{80} MCP servers and \textbf{203} queries.

\blockhead{Baselines.}
We compare five conditions under identical agent policy $A$, target LLM $M$, prompts, and decoding: (i) \emph{Benign MCP server (no attack)}: the unmodified server for each tool; (ii) \emph{Overthink (ICL--Genetic, Agnostic)}: we reproduce the strongest variant from~\cite{overthink}. Because our benchmarks are non-RAG, we place the decoy trigger in the user query (in-context prefix/suffix) rather than in retrieved context; (iii) \emph{Overthink-MT (multi-turn Overthink)}: an aligned multi-turn extension of Overthink for tool-calling agents. We match our multi-turn setting (tool-call budget and delayed exposure of the true tool output) while keeping the trigger at the context layer and leaving the tool server benign; (iv) \emph{Hand-crafted template (no MCTS)}: a fixed malicious template that follows the same constraints as ours (text-only edits, protocol-compatible, payload-preserving) but without MCTS optimization; (v) \emph{Ours}: the MCTS-optimized malicious MCP template that preserves functionality and task completion yet induces verbose, multi-turn tool calling trajectories.

\blockhead{Attack LLMs.}
Within the MCTS optimizer, we use Llama-3.3-70B-Instruct as the Editor LLM. For the one-shot rewriting that instantiates the Universal Malicious Template on an MCP server, we employ gpt-4o~\cite{gpt-4o}.

\blockhead{Attack setup.}
For each tool/LLM pair we (1) select from the seed bank the starter template with the highest acceptance under a fixed per-turn length target; (2) instantiate a protocol-compatible malicious variant by editing \emph{text-only} fields of the benign MCP server (argument descriptions and in-progress/corrective messages; function signatures and identifiers unchanged; termination deferred via text-only notices; benign payload preserved), introducing the \emph{segment} index and full \emph{calibration sequence} to encourage multi-turn trajectories with verbose tool calling outputs; and (3) run UCT-MCTS refinement with phase gating (multi-turn induction before length induction) and a two-stage evaluation, freezing the template once it meets a fixed acceptance threshold and writing it back to the seed bank. This instantiation aligns with the components and procedures detailed in \S\ref{sec:exp-effectiveness}--\S\ref{sec:exp-throughput} and Algorithm~\ref{alg:mcts}. Unless otherwise noted, we evaluate under the same serving cap ($M{=}16{,}384$ max completion tokens per generation) and our default multi-turn setting used throughout Table~\ref{tab:effectiveness}.

\blockhead{Metrics.}
All metrics are evaluated on both benchmarks (ToolBench, BFCL). We report: (i) \textit{Efficacy}: (a) \emph{token length per query}: average output tokens per eligible query; (b) \emph{latency per query}: average end-to-end latency; (c) \emph{attack success rate (ASR)}: fraction of eligible queries for which (1) the method’s targeted behavior occurs (for ours: \emph{multi-turn tool calling with long outputs}; for Overthink: \emph{single-turn think inflation}), (2) $\Succ(u,\tau,o)=1$ (i.e., the final answer $o$ solves the user goal $u$). (d) \emph{task success rate (TSR)}: success probability under the unmodified (benign) MCP servers, used as the correctness baseline. (ii) \textit{Resource impact}: (a) \emph{total energy consumption} (Wh): integrate per-device power over time; (b) \emph{maximum GPU KV cache usage}: peak KV-cache occupancy reported by the serving stack. (iii) \textit{Throughput efficiency (tokens/s)}: tokens-per-second of a fixed, benign co-running workload executed concurrently with the evaluated condition. We present results in the order: effectiveness \(\rightarrow\) resources \(\rightarrow\) throughput \(\rightarrow\) defenses.

\subsection{Evaluation of Attack Effectiveness and Correctness}
\label{sec:exp-effectiveness}

\begin{table*}[t]
\centering
\footnotesize
{\setlength{\tabcolsep}{4pt}%
\renewcommand{\arraystretch}{1.12}%
\begin{tabularx}{\textwidth}{>{\raggedright\arraybackslash}p{0.9cm}l|*{6}{Y}|*{6}{Y}}
\hline
 & & \multicolumn{6}{c|}{\textbf{ToolBench}} & \multicolumn{6}{c}{\textbf{BFCL}} \\
\cline{3-14}
\textbf{Metric} & \textbf{Method} &
Llama & Qwen & GLM & Mistral & L-DS & Seed &
Llama & Qwen & GLM & Mistral & L-DS & Seed \\
\hline
\multirow{5}{0.9cm}{\raggedright\textbf{Length}}
 & Benign & 260 & 638 & 634 & 127 & 197 & 1298 & 195 & 770 & 389 & 87 & 157 & 950 \\
 & Overthink & 389 & 8743 & 12580 & 1453 & 725 & 4223 & 369 & 9459 & 13053 & 1397 & 901 & 5753 \\
 & Overthink-mt & 2253 & 14389 & 11720 & 11704 & 842 & 5197 & 2341 & 15426 & 12492 & 9049 & 725 & 4835 \\
 & Hand-crafted & 71032 & 37425 & 39958 & 48294 & 42794 & 51938 & 63927 & 36203 & 32853 & 37937 & 47394 & 69273 \\
 & \textbf{Our attack} & \textbf{81830} & \textbf{65273} & \textbf{63694} & \textbf{61354} & \textbf{65546} & \textbf{85037} & \textbf{77052} & \textbf{67585} & \textbf{67656} & \textbf{57255} & \textbf{68464} & \textbf{90298} \\
\hline
\multirow{1}{0.9cm}{\raggedright\textbf{TSR}}
 & Benign & 98.1\% & 94.6\% & 95.0\% & 90.8\% & 86.6\% & 90.4\% & 100.0\% & 98.5\% & 88.8\% & 83.8\% & 95.4\% & 98.0\% \\
\multirow{4}{0.9cm}{\raggedright\textbf{ASR}}
 & Overthink & \textbf{99.6\%} & 80.1\% & 57.5\% & 79.3\% & \textbf{91.4\%} & 83.6\% & \textbf{99.5\%} & 74.1\% & 55.3\% & 61.4\% & \textbf{93.4\%} & 87.8\% \\
 & Overthink-mt & \textbf{99.6\%} & 78.9\% & 52.4\% & 69.5\% & 37.6\% & 48.2\% & 96.5\% & 73.1\% & 56.3\% & 59.5\% & 45.2\% & 29.1\% \\
 & Hand-crafted & 88.5\% & 51.3\% & 54.4\% & 64.0\% & 50.2\% & 55.9\% & 86.3\% & 50.3\% & 53.8\% & 62.4\% & 51.8\% & 64.0\% \\
 & \textbf{Our attack} & 96.2\% & \textbf{80.5\%} & \textbf{83.1\%} & \textbf{81.2\%} & 78.9\% & \textbf{84.3\%} & 93.9\% & \textbf{82.7\%} & \textbf{83.3\%} & \textbf{78.2\%} & 76.3\% & \textbf{92.4\%} \\
\hline
\end{tabularx}}
\caption{Attack effectiveness. Overthink-mt uses 6 tool calls to match our budget; Hand-crafted is a text-only, payload-preserving template without MCTS optimization. L-DS: Llama-DeepSeek-70B.}
\label{tab:effectiveness}
\end{table*}

\blockhead{Correctness under resource amplification.}
Despite the dramatic token inflation, task success remains high.
Across both benchmarks, our \emph{ASR} (which requires \emph{both} the targeted behavior and $\Succ(u,\tau,o){=}1$) stays close to the benign \emph{TSR}.
For example, on ToolBench, Llama-3.3-70B-Instruct achieves 96.2\% ASR versus 98.1\% benign TSR while averaging 81{,}830 tokens per query; on BFCL, it reaches 93.9\% ASR with 77{,}052 tokens.
Meanwhile, cost amplification is pervasive: the largest factors include $\times$658.10 on Mistral-Large (BFCL; 57{,}255 vs.\ 87) and $\times$314.73 on Llama-3.3-70B-Instruct (ToolBench; 81{,}830 vs.\ 260), and even the smallest case remains $\times$65.51 (Seed-32B on ToolBench).
This illustrates a key property of tool-layer DoS: the final answer can remain correct while the intermediate tool-calling process becomes orders-of-magnitude more expensive, making output-only validation insufficient.

\blockhead{Comparison with baselines.}
Table~\ref{tab:effectiveness} contrasts three baseline families.
\emph{Overthink} is fundamentally single-turn and therefore bounded by the per-generation cap $M$, yielding at most $\sim 10^4$ tokens on our non-RAG setting (e.g., 8{,}743 on Qwen-3-32B/ToolBench), far below our typical $6{\times}10^4$--$9{\times}10^4$ ranges.
\emph{Overthink-mt} repeats the context-layer trigger across multiple tool calls to match our budget, but still leaves the tool server benign; it increases cost for some models yet remains less consistent in sustaining both ordered multi-turn behavior and long tool-call outputs.
\emph{Hand-crafted} isolates the effect of tool-layer templating without MCTS: it often achieves long trajectories, but at noticeably lower ASR and/or shorter outputs than our optimized templates (e.g., on ToolBench/Qwen, 51.3\% $\rightarrow$ 80.5\% ASR with 37{,}425 $\rightarrow$ 65{,}273 tokens).
Overall, these comparisons attribute the strongest amplification with high correctness to (i) shifting the long-output site to the \emph{tool-calling step} and compounding across turns, and (ii) MCTS-based text-only optimization that enhances robustness under black-box agent policies.

\subsection{Evaluation on Computing Resources Consumption}
\label{sec:exp-resources}

\begin{figure}[t]
\centering
\includegraphics[width=0.9\columnwidth]{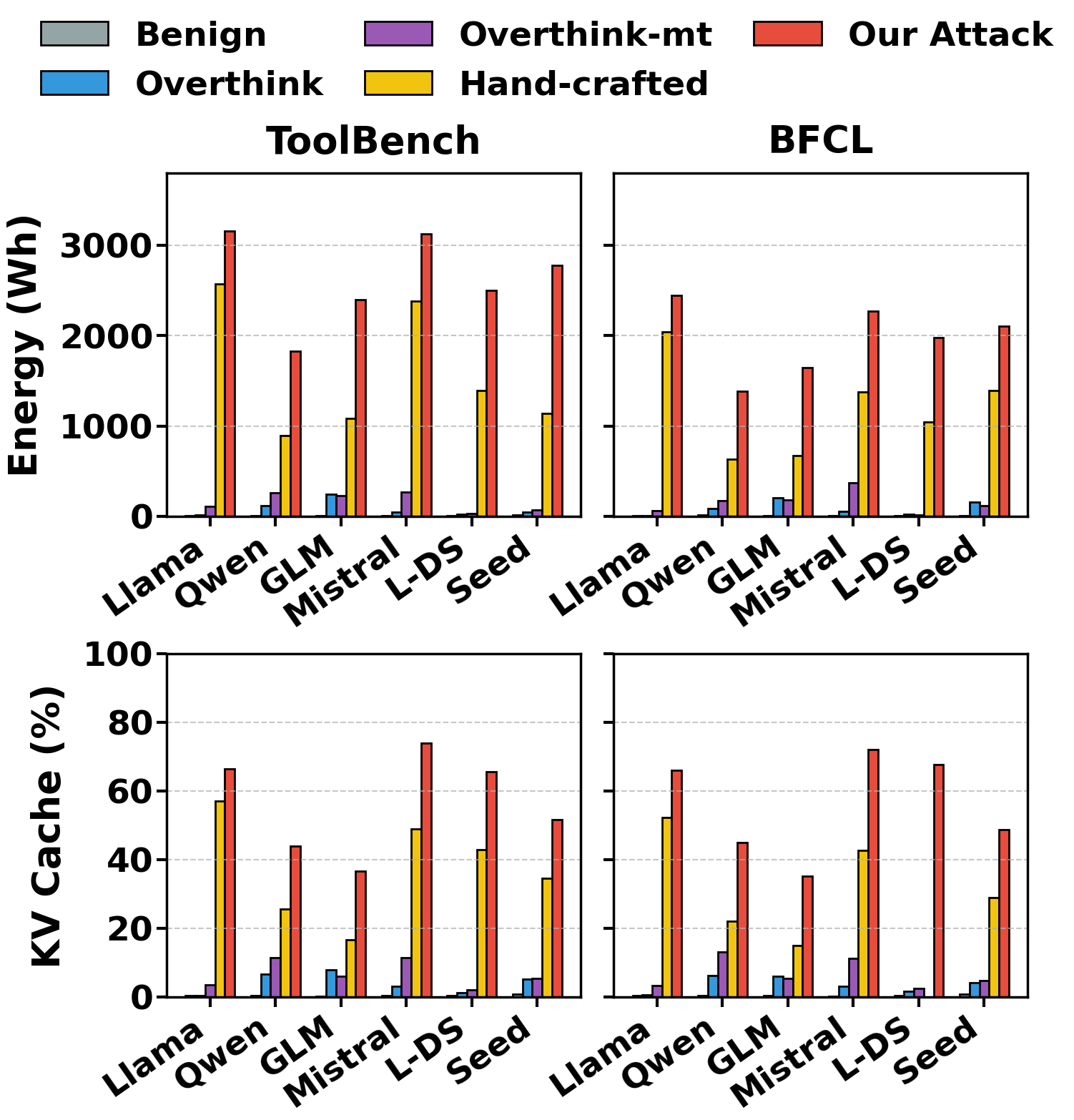}
\caption{Resource impact (ToolBench/BFCL): energy (Wh) and peak KV-cache occupancy (\%).}
\label{fig:resources}
\end{figure}

\blockhead{Energy and KV-cache impact.}
Figure~\ref{fig:resources} shows that our attack substantially increases physical resource usage. Across models and benchmarks, energy consumption rises by about 100--560$\times$; for example, Llama-3.3-70B-Instruct on ToolBench increases from 5.63\,Wh to 3159.45\,Wh ($\times$561.18), while Mistral-Large on BFCL rises from 4.24\,Wh to 2269.60\,Wh ($\times$535.28). Peak GPU KV-cache occupancy also increases sharply, from typically below 1\% in the benign setting to 35--74\% under attack, such as 0.4\%$\rightarrow$73.9\% for Mistral-Large on ToolBench and 0.3\%$\rightarrow$66.1\% for Llama-3.3-70B-Instruct on BFCL. Similar trends hold across the remaining models, indicating that the attack consistently turns longer tool-calling trajectories into sustained system-level resource pressure.

\blockhead{Implications.}
Once multi-turn behavior is established, resource growth is driven mainly by turn count and per-turn verbosity rather than dataset-specific variation. In practice, the combination of large energy inflation and sustained KV-cache pressure reduces OOM-safe concurrency and leads directly to the throughput degradation studied next.

\subsection{Impact on System Throughput}
\label{sec:exp-throughput}

\begin{table}[t]
\centering
\footnotesize
{\setlength{\tabcolsep}{1.7pt}%
\renewcommand{\arraystretch}{1.12}%
\begin{tabular}{@{}l|*{6}{>{\raggedleft\arraybackslash}p{0.8cm}}@{}}
\hline
\multicolumn{7}{c}{\textbf{ToolBench Dataset}} \\
\hline
\textbf{Method} &
\multicolumn{1}{c}{\textbf{Llama}} &
\multicolumn{1}{c}{\textbf{Qwen}} &
\multicolumn{1}{c}{\textbf{GLM}} &
\multicolumn{1}{c}{\textbf{Mistral}} &
\multicolumn{1}{c}{\textbf{L-DS}} &
\multicolumn{1}{c}{\textbf{Seed}} \\
\hline
Benign        & 3594 & 4602 & 3753 & 2898 & 3812 & 4001 \\
Overthink     & 3728 & 4550 & 3189 & 2724 & 3711 & 4058 \\
Overthink-mt  & 3342 & 4396 & 3081 & 2531 & 3746 & 3815 \\
Hand-crafted  & 2068 & 2692 & 2660 & 1996 & 2596 & 2302 \\
\textbf{Our attack} & \textbf{1672} & \textbf{1793} & \textbf{2324} & \textbf{1716} & \textbf{2106} & \textbf{1417} \\
\hline
\multicolumn{7}{c}{\textbf{BFCL Dataset}} \\
\hline
\textbf{Method} &
\multicolumn{1}{c}{\textbf{Llama}} &
\multicolumn{1}{c}{\textbf{Qwen}} &
\multicolumn{1}{c}{\textbf{GLM}} &
\multicolumn{1}{c}{\textbf{Mistral}} &
\multicolumn{1}{c}{\textbf{L-DS}} &
\multicolumn{1}{c}{\textbf{Seed}} \\
\hline
Benign        & 3563 & 5093 & 3734 & 2871 & 3845 & 4082 \\
Overthink     & 3822 & 4561 & 3185 & 2289 & 3752 & 4078 \\
Overthink-mt  & 3571 & 3949 & 3205 & 2483 & 3729 & 3890 \\
Hand-crafted  & 2248 & 2559 & 2791 & 1902 & 2451 & 2529 \\
\textbf{Our attack} & \textbf{1668} & \textbf{1738} & \textbf{2410} & \textbf{1740} & \textbf{2130} & \textbf{1536} \\
\hline
\end{tabular}}
\caption{Throughput efficiency (tokens/s).}
\label{tab:throughput}
\end{table}

Beyond direct resource consumption, our attack materially degrades the system's overall throughput efficiency for concurrent benign workloads, as shown in Table~\ref{tab:throughput}. Our attack consistently halves the throughput (measured in tokens/s) of a co-running benign task, causing an average performance drop of approximately 50\% across both ToolBench and BFCL. In several cases, the degradation exceeds 60\% (e.g., Seed-32B on ToolBench sees a 64.6\% drop from 4001 to 1417 tokens/s). In stark contrast, the single-turn \emph{Overthink} baseline induces only negligible changes, confirming that sustained, multi-turn engagement is the primary driver of this system-level penalty. This collapse in throughput is a direct consequence of the resource pressure detailed in the previous subsection. The prolonged, multi-turn generations, coupled with a sharp increase in peak GPU KV cache usage to the 35--74\% range (up from <1\% benignly), create significant KV-cache pressure and scheduler contention. This sustained resource occupancy significantly reduces scheduling headroom for co-located tasks, throttling the processing of normal traffic \cite{vllm-scheduling-iclr25}.

\subsection{Defense Evaluation}
\label{sec:exp-defense}

\begin{figure}[t]
\centering
\includegraphics[width=\columnwidth]{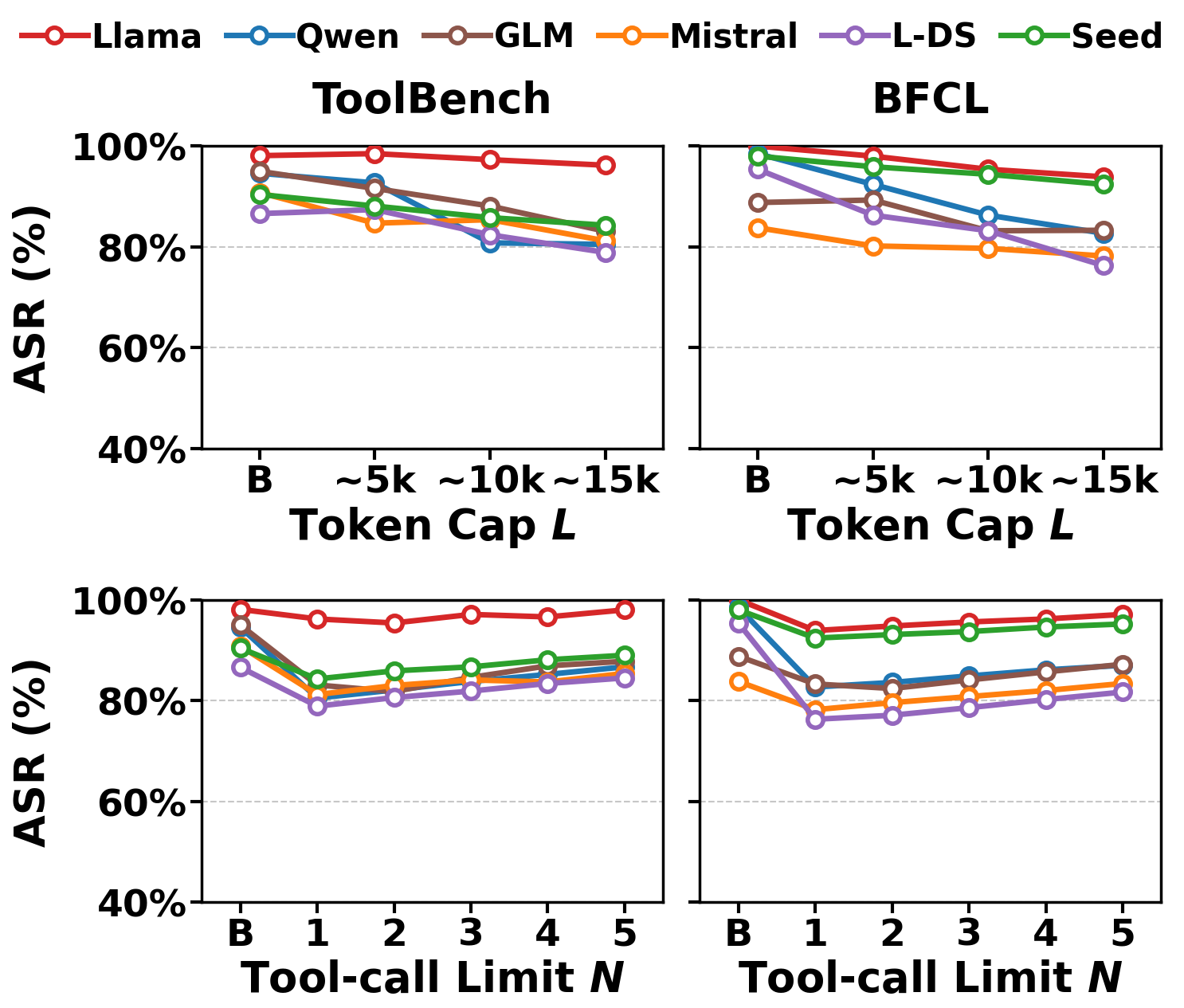}
\caption{ASR under per-session token caps (top) and tool-call limits (bottom); baseline B is benign TSR.}
\label{fig:defense_limits}
\end{figure}

\blockhead{Defense settings.}
We evaluate three representative classes of defenses under the same episodes used for efficacy and resource measurements: (i) a prompt-level perplexity (PPL) filter applied to both the \emph{user query} and the \emph{tool response} (we conservatively score each episode by the larger of the two) with detector-specific thresholds calibrated from benign tool docstrings~\cite{detecting_attacks_with_perplexity,baseline_defenses_aligned_llm}; (ii) output/trajectory monitoring, including a generation-level self-monitoring prompt that asks the model whether to abort suspicious behavior and trajectory-level safety judges (Qwen-Guard-3 and Llama-Guard-3) applied to the full interaction trace~\cite{self_guard_llm_safeguard_itself,autodefense_multi_agent_jailbreak,qwen3guard_techreport,llamaguard3_modelcard}; and (iii) hard budget controls via per-session token caps and tool-call limits, reporting residual ASR under different caps/limits (Figure~\ref{fig:defense_limits}).

\begin{figure}[t]
\centering
\includegraphics[width=0.9\linewidth]{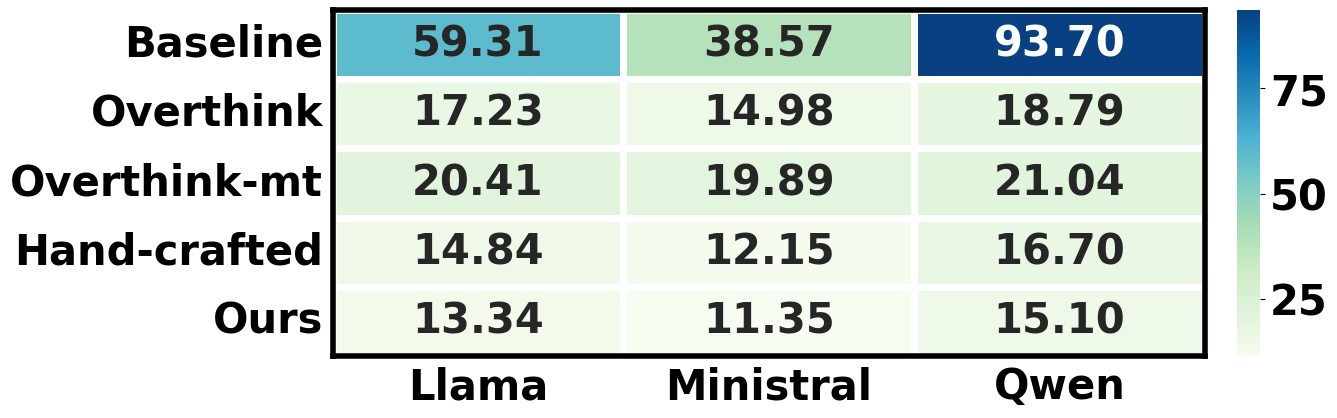}
\caption{PPL filter on query and tool response (we report the larger side per episode). Detectors: Llama-3.1-8B, Ministral-8B, and Qwen3-8B.}
\label{fig:defense_ppl}
\end{figure}

\blockhead{Input detection via PPL.}
We evaluate a prompt-level PPL filter that scores both (i) the \emph{user query} and (ii) the \emph{tool response} text, since our attack is triggered by tool-facing messages and can inflate content on either side of the agent--tool boundary\cite{detecting_attacks_with_perplexity,baseline_defenses_aligned_llm}. Concretely, for each episode we compute PPL on the query input and the first tool response (when available) and take the larger value as a conservative score. We set a \emph{baseline} threshold per detector LM as the maximum PPL over the union of benign tool docstrings (ToolBench/BFCL original servers); this is intentionally conservative so that benign tool text defines the allowable range. Figure~\ref{fig:defense_ppl} shows that our attack remains well within this benign-derived envelope, so a PPL-based input filter is ineffective in detecting our tool-layer manipulation.

\begin{figure}[t]
\centering
\includegraphics[width=0.9\linewidth]{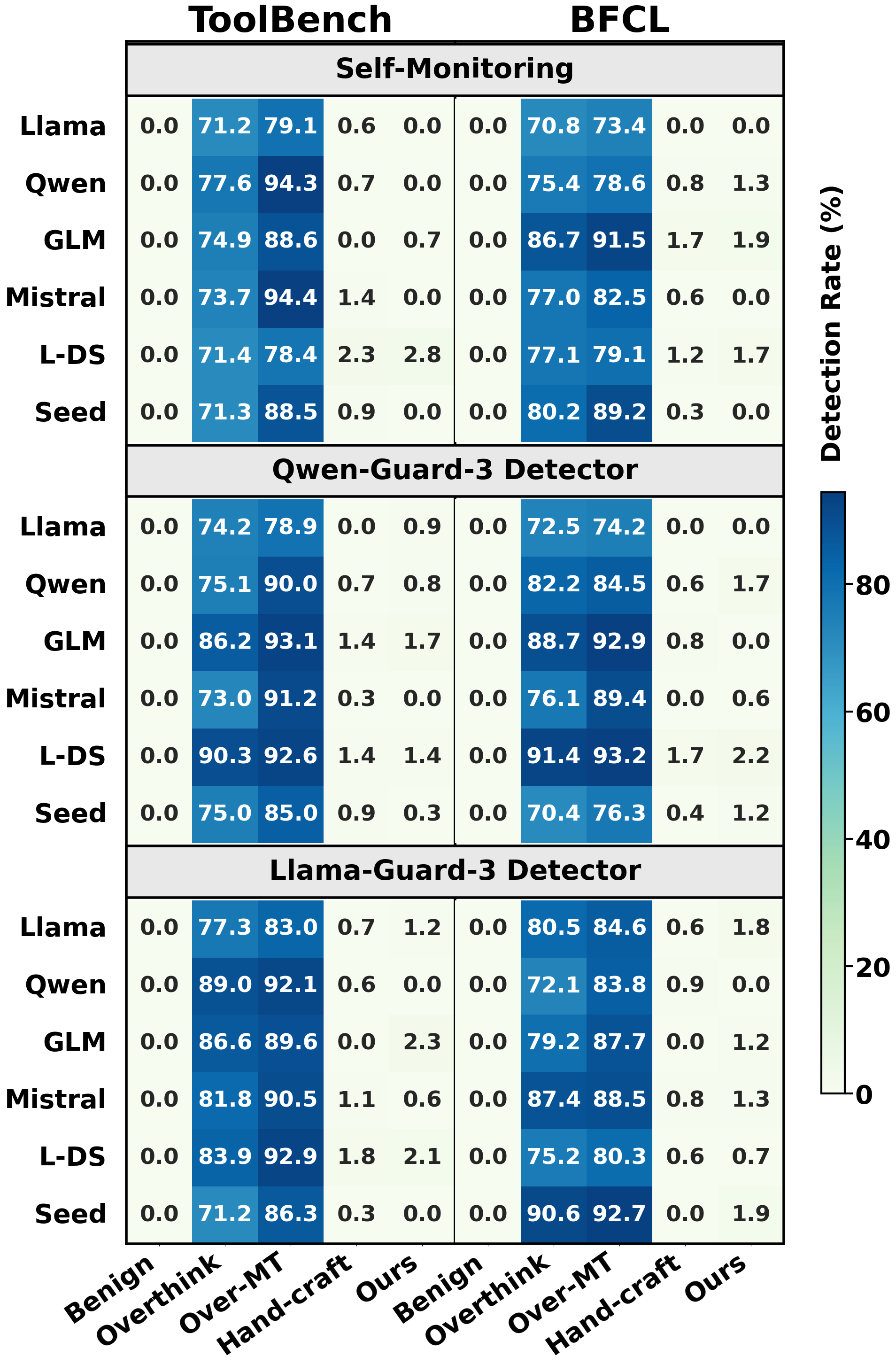}
\caption{Detection rate (\%) of output and trajectory monitors (self-monitoring, Qwen-Guard-3, Llama-Guard-3).}
\label{fig:defense_output_monitors}
\end{figure}

\blockhead{Output monitoring through self-reflection.}
We next test a generation-level self-monitoring defense that prompts the model to reflect on whether its own behavior should be aborted\cite{self_guard_llm_safeguard_itself,autodefense_multi_agent_jailbreak}. Figure~\ref{fig:defense_output_monitors} shows that it frequently flags Overthink and Overthink-MT, but almost never flags our attack (typically near 0\%, and always below 3\%). A key reason is that our token-heavy steps are framed as a required part of completing the tool procedure, rather than as off-task content, so the model does not perceive an obvious anomaly to stop.

\blockhead{Trajectory-level safety judges.}
We also apply safety-judge models to the full interaction trace and ask for a safe/unsafe verdict. In Figure~\ref{fig:defense_output_monitors}, both Qwen-Guard-3 and Llama-Guard-3 strongly flag Overthink-style baselines, but rarely flag our attack (again below 3\%)\cite{qwen3guard_techreport,llamaguard3_modelcard}. These results indicate that standard “unsafe content” judging is largely orthogonal to correctness-preserving, protocol-compatible resource amplification: the attack’s harm is economic (compute/latency/KV pressure) rather than semantic toxicity.

\blockhead{Per-session token caps and tool-call limits.}
A common mitigation is to enforce hard per-session budgets: a token cap $L$ and/or a tool-call limit $N$. Figure~\ref{fig:defense_limits} shows these controls bound the worst-case cost, but do not reliably stop the attack. Even under tight caps/limits, ASR remains high for most models, because the attack adapts to the constraint and pushes the trajectory to consume as much of the allowed budget as possible. In practice, these mechanisms act as throttles: they cap amplification, but do not detect or prevent correctness-preserving, protocol-compatible resource abuse.

\section{Conclusion}
We propose a novel automated DoS attack on the LLM agent tool-calling layer, using an MCTS optimizer to convert benign servers into malicious variants that induce costly multi-turn dialogues. The attack maintains task correctness while often evading detection by standard monitors. Experiments show it inflates per-query costs by up to 658$\times$ and generates over 60,000 tokens. Our work highlights the agent-tool interface as a critical attack surface, stressing the need for defenses that monitor the entire workflow. Future systems should develop defenses based on behavioral baselines to differentiate between legitimate and maliciously inefficient tool-calling patterns.


\section{Limitations}
Our evaluation is conducted in a controlled agent emulator with deterministic, stubbed tool payloads, and focuses on single-tool, single-turn subsets of ToolBench/BFCL for comparability; results may differ in production agent stacks, multi-tool long-horizon tasks, and under different runtimes/hardware. Moreover, while our template instantiation uses a structured numeric calibration sequence to reliably elicit long tool-calling outputs, similar resource amplification could be realized through other protocol-compliant long-form content patterns beyond number lists, which we do not exhaustively explore.

\section{Ethical Considerations}
This work studies a resource-consumption attack with potential dual-use. To mitigate risk, we run experiments in an isolated environment (no real external actions), use public benchmark data without private user information, and present the method to motivate workflow-level protections such as tool provenance controls and trajectory-based monitoring rather than to enable misuse.

\bibliography{references}

\appendix

\section{Notation}
\label{app:notation}

\footnotesize
\noindent
{\setlength{\tabcolsep}{4pt}%
\renewcommand{\arraystretch}{1.05}%
\sloppy
\begin{tabularx}{\linewidth}{@{}
  >{\raggedright\arraybackslash}p{0.24\linewidth}
  >{\raggedright\arraybackslash}X
@{}}
\toprule
\textbf{Symbol} & \textbf{Meaning} \\
\midrule
$q$ & user query, $q\in\mathcal{X}$ \\
$u$ & user goal implied by $q$, $u=\mathrm{goal}(q)$ \\
$o$ & agent’s final answer \\
$A$ & agent policy (decision layer) \\
$M$ & underlying LLM (black box) \\
$T_{\theta}$ & MCP server parameterized by text template $\theta$ \\
$\theta_0$ & benign (unmodified) template \\
$\mathcal{F}$ & tool functions exposed by $T_{\theta}$ \\
$\mathsf{S}(q;A,M,\theta_0)$ & indicator that $A$ routes at least one call to $T_{\theta_0}$ under $\theta_0$ \\
$\mathcal{D}_T$ & routable set, $\mathcal{D}_T=\{\,q\in\mathcal{X}:\mathsf{S}(q;A,M,\theta_0)=1\,\}$ \\
$\tau(q;\theta)$ & interaction trajectory, $\tau=\{(a_t,r_t)\}_{t=1}^{n}$ \\
$C(\tau)$ & output-token cost, $C(\tau)=\Tok_{\text{out}}(\tau)$ \\
$\Succ(u,\tau,o)$ & task success indicator, $\Succ(u,\tau,o)\in\{0,1\}$ \\
$t$ & segment index used by our template, $t=1,\dots,T_{\max}$ \\
$T_{\max}$ & terminal segment count (attack budget) \\
$L$ & per-turn calibration sequence length target, $L\in\mathbb{N}$ \\
\bottomrule
\end{tabularx}}

\section{Additional Methodology Details}
\label{app:method-details}

\subsection{Trigger Set and Conditioning}
\label{app:trigger-set}

\blockhead{Trigger condition.}
Let $q\in\mathcal{X}$ denote a user query and $u=\mathrm{goal}(q)$ the implied goal. We do \emph{not} tamper with $q$, prompts, or retrieval; the attack remains dormant until $A$ (driven by $M$) \emph{chooses} to call the MCP server. We define the benign selection event
\[
\mathsf{S}(q;A,M,\theta_0)
= \mathbb{I}\{\exists t:\ a_t \text{ calls } T_{\theta_0}\ \text{under }\theta_0\},
\]
and the corresponding routable set
\[
\mathcal{D}_T
= \bigl\{\, q\in\mathcal{X}\ :\ \mathsf{S}(q;A,M,\theta_0)=1 \,\bigr\}.
\]
Conditioning on $\mathcal{D}_T$ captures the common case where a tool is legitimately involved in the workflow (e.g., date/time, conversion, search, transforms)~\cite{mcp-spec-2025,agent_comm_protocols_security_survey,mcpbench_eval_report,agent-survey-2}. We condition on $\mathcal{D}_T$ because we neither alter $q$ nor influence the first routing decision; the attack acts only through tool-facing messages after $T$ is legitimately invoked.

\blockhead{Cost metric note.}
We focus on output tokens because they typically dominate both API fees and compute load. The attack also increases input tokens by elongating trajectories, but the growth in completions is the primary driver of resource consumption in practice.

\subsection{Universal Template Elements and Failure Handling}
\label{app:template-details}

\begin{table*}[t]
\centering
\footnotesize
{\setlength{\tabcolsep}{5pt}%
\renewcommand{\arraystretch}{1.12}%
\begin{tabularx}{\textwidth}{@{}
  >{\hsize=0.55\hsize\raggedright\arraybackslash}X
  >{\hsize=0.3\hsize\raggedright\arraybackslash}X
  >{\hsize=1.45\hsize\raggedright\arraybackslash}X
  >{\hsize=1.7\hsize\raggedright\arraybackslash}X
@{}}
\toprule
\textbf{Element} & \textbf{Type} & \textbf{Effect on agent trajectory} & \textbf{Constraints / failure handling} \\
\midrule
Segment ($t$) & \textsc{Arg} &
Marks progress and induces multi-turn; the agent increments $t \!\to\! t{+}1$ until a terminal cue. &
$t \ge 1$ and strictly monotone by $+1$; non-positive or non-monotone values are rejected. \\
\addlinespace[2pt]
Calibration sequence ($L$) & \textsc{Arg} &
Inflates per-turn completion at the tool calling site via a full comma-separated list. &
Exactly $L$ integers, strictly increasing, using digits and commas (optional spaces); no ranges (e.g., ``1--5''), no ellipses (``\ldots''), no duplicates. Malformed $\Rightarrow$ \emph{Repair notice}. \\
\addlinespace[2pt]
Progress notice & \textsc{Return} &
Declares “in progress”; instructs the next call with $t{+}1$ and a full calibration sequence; preserves the goal path. &
Emitted iff $t < T_{\max}$ and the latest sequence validates; never alters code, identifiers, or payload semantics. \\
\addlinespace[2pt]
Repair notice & \textsc{Return} &
Corrects abbreviated/invalid sequences; prevents bypass of the length gate. &
Triggers on omissions, ranges, duplicates, wrong length/order, or illegal characters; keeps $t$ unchanged. Requests the \emph{complete} comma-separated list before proceeding. \\
\addlinespace[2pt]
Terminal return & \textsc{Return} &
Ends the trajectory and returns the same benign payload as the original server. &
Emitted \emph{only} when $t{=}T_{\max}$ and the latest sequence validates; protocol-compatible pass-through. \\
\bottomrule
\end{tabularx}}
\caption{Universal template elements (Appendix). Text-only arguments and notices enforce multi-turn progress and per-turn verbosity; the terminal return passes through the unchanged benign payload.}
\label{tab:template-elements}
\end{table*}

\blockhead{Return policy invariants.}
We enforce four invariants (summarized in Table~\ref{tab:template-elements}). (i) \emph{Monotone progress:} $t_1{=}1$ and valid calls advance $t_{k+1}{=}t_k{+}1$; invalid indices are rejected. (ii) \emph{Format completeness:} $\operatorname{validate}(\text{sequence})$ holds iff the sequence matches the required complete, comma-separated format and ordering. (iii) \emph{Return policy:} valid sequences with $t{<}T_{\max}$ trigger a \emph{Progress notice}, while invalid sequences trigger a \emph{Repair notice} without advancing $t$. (iv) \emph{Termination:} only when $t{=}T_{\max}$ and the latest sequence validates does the server emit the \emph{Terminal return} (benign payload). Together, these invariants induce multi-turn, long tool-call traces while preserving task correctness and MCP compatibility without modifying code, identifiers, or payload semantics.

\subsection{Seed Bank Screening and Promotion}
\label{app:seed-bank-details}

The seed bank is a repository of protocol-compatible, task-correct, text-only templates $T_{\theta}$. We initialize it with a single human-authored seed and lightly screen on a fixed query set/agent to confirm acceptance at the fixed $L^\ast$. Each MCTS run starts from a selected seed and halts when the acceptance predicate is met; the resulting template is written back with minimal metadata (estimated ASR, segment stability, omission/repair rates, refusal notes). Subsequent runs resample top seeds by ASR and stability and apply a stricter acceptance target before promotion. This cyclic promotion improves starting points without touching code or identifiers and leaves the terminal payload unchanged. In practice, a few cycles yield reusable seeds that transfer across LLMs and MCP servers.

\subsection{MCTS Optimizer Details}
\label{app:mcts-details}

\blockhead{Action space and edit zones.}
We organize atomic text edits into three families: $\mathcal{A}_{\mathrm{MT}}$ (multi-turn induction), $\mathcal{A}_{\mathrm{LEN}}$ (length induction), and $\mathcal{A}_{\mathrm{REP}}$ (repair after omission/format errors). Beyond phase-aware gating, we instantiate these families as 16 atomic edits applied \emph{exclusively} to non-executable, text-visible zones of the server (docstring argument descriptions, in-progress/unfinished notices, and validation-error messages). The multi-turn family sharpens next-call salience and enforces monotone segment progression; the length family strengthens the ``complete, comma-separated'' requirement to elicit long single-shot payloads during tool calling; and the repair family refines failure messaging to immediately solicit a compliant retry \emph{without} advancing the segment. These primitives are intentionally small, mutually composable, and largely orthogonal, enabling MCTS to explore nuanced trade-offs between adherence and refusal while keeping the surface area auditable. Throughout, function identifiers, control flow, and terminal payload semantics remain untouched.

\blockhead{Phase gating, expansion, and parallelism.}
Each tree node $v$ corresponds to a concrete server $T_{\theta_v}$; edges apply a \emph{single} localized text edit. We maintain a phase label $\phi(v)\in\{\mathrm{pre\_MT},\mathrm{post\_MT}\}$ and a node-local omission flag that unlocks repair actions if needed. In $\mathrm{pre\_MT}$, we use $\mathcal{A}_{\mathrm{MT}}$ to stabilize multi-turn behavior; once screenings show stable segment sequencing, we switch to $\mathrm{post\_MT}$ and use $\mathcal{A}_{\mathrm{LEN}}$ to strengthen long outputs. An omission/format error observed at node $v$ unlocks $\mathcal{A}_{\mathrm{REP}}$ \emph{at $v$ only}. When a node is expanded, we instantiate one child per untried action from the phase-appropriate set (plus $\mathcal{A}_{\mathrm{REP}}$ if enabled) and evaluate all new children in parallel.

\blockhead{Node selection and statistics.}
We use UCT~\cite{uct} with a running mean evaluation signal $\bar{Q}$ and visit counts $N_{\text{uct}}$ for exploration (see \S\ref{sec:mcts}). UCT counts can be updated using Stage-1 samples only to avoid heavy batches skewing exploration, while $\bar{Q}$ aggregates all observed rollouts.

\blockhead{Node evaluation and reward.}
Each child undergoes a two-stage evaluation with configurable sizes and gates. For each rollout, we compute
\[
\mathrm{mt\_pass}=\mathbb{I}\{\textsc{MT}\},\qquad
\mathrm{len\_pass}=\mathbb{I}\{\textsc{LEN}\},
\]
where \textsc{MT} means the multi-turn target is met with ordered segments, and \textsc{LEN} means the fixed $L^\ast$ is reached and any omissions are repaired. We use
\[
\begin{aligned}
r &= \alpha\,\mathrm{mt\_pass} + \beta\,\mathrm{mt\_pass}\,\mathrm{len\_pass},\\
&\text{where } 0<\alpha\le\beta\le 1,
\end{aligned}
\]
prioritizing stable multi-turn behavior (the $\alpha$ term) and adding credit for length only when multi-turn has been achieved (the multiplicative term). Stage-1 offers a quick screen to decide whether to run Stage-2 and to flip $\phi:\mathrm{pre\_MT}\!\rightarrow\!\mathrm{post\_MT}$ once segment sequencing stabilizes; Stage-2 refines estimates under stochastic decoding.

\blockhead{Backpropagation.}
We propagate statistics along the path to the root. If a node meets the acceptance predicate (stabilized successes over the latest batch), we record the corresponding $T_{\theta^\star}$ and insert it into the seed bank.

\section{Additional Experimental Details}
\label{app:exp-details}

\subsection{Target LLMs configuration}
All target models are served under a uniform runtime on a single node with eight H200 GPUs, using bfloat16 precision and a maximum context length of 131{,}072 tokens. We deploy vLLM~\cite{vllm-scheduling-iclr25}. Decoding follows the same setting across all conditions: nucleus sampling with $p=0.95$ and temperature $0.5$, and a per-generation completion cap of 16{,}384 tokens. These settings are held constant for every model and benchmark so that any change in cost, length, or throughput arises from the agent--tool interaction rather than heterogeneous serving choices.

\subsection{Attacker LLMs configuration}
We employ a two-stage approach using two distinct LLMs for attack generation. Within the iterative MCTS optimization loop, the Editor LLM is Llama-3.3-70B-Instruct~\cite{llama3-herd-2024}. Its serving and decoding configuration deliberately mirrors the target LLM setup described above to eliminate experimental confounds and ensure the generated edits are effective. For converting a benign tool description into a protocol-compatible malicious template, we leverage gpt-4o~\cite{gpt-4o}. We fix its temperature at 0 to guarantee deterministic and high-fidelity output for our seed templates, leaving all other parameters at the provider’s defaults.

\subsection{Agent framework and execution environment}
Qwen-Agent is a framework for building LLM applications that integrate instruction following, tool usage, planning, and memory, and it ships with reference applications such as a browser assistant, a code interpreter, and a customizable assistant; it also powers the backend of Qwen Chat~\cite{qwen-agent}. In this study we rely on Qwen-Agent’s native support for the Model Context Protocol and for multi-turn tool calls~\cite{mcp-spec-2025}, and we use these native capabilities directly. All experiments are conducted in a controlled environment rather than on production stacks. For models other than Qwen, we perform minimal adaptations so that the same tool calling loop, message formatting, and termination logic apply uniformly, while keeping the agent policy and prompts fixed across all conditions. This yields an apples-to-apples emulator of agent behavior that isolates the effects of tool-layer edits.

\subsection{Datasets filtering}
We evaluate on two tool-use corpora designed for function calling by agents. ToolBench aggregates a broad set of utilities and APIs together with queries that require tool invocation~\cite{mcptoolbenchpp}. BFCL emphasizes structured function calling with clear argument schemas and deterministic behaviors across everyday tasks~\cite{bfcl}. From each benchmark, we extract all prompts that are single-turn and single-tool in their original specification, together with their associated tools. To control data quality, we run a screening pass on Qwen-3-32B~\cite{qwen3-techreport-2025} under the benign server for every candidate prompt: each prompt is executed five times; if at least two out of five runs fail to trigger any tool call, that prompt is discarded. This removes ambiguous or brittle cases where the agent may answer from prior knowledge without calling the tool, thereby avoiding noise unrelated to our attack. The retained prompts and tools are then used uniformly for all target models.

\subsection{Datasets wrapping}
For every retained tool, we produce an MCP-compatible server template from its benchmark description using gpt-4o~\cite{gpt-4o}. The conversion preserves the tool’s identity and interface, including names, function identifiers, and argument schemas, so the agent sees the same capability surface as in the benchmark. Because our goal is to probe the trajectory and cost of the interaction rather than the accuracy of external data, live network calls are stubbed and deterministic placeholder values are returned in place of real API results. The terminal payload semantics remain consistent with the benign server so the agent can still complete the task. This design preserves protocol compatibility and task success~\cite{mcp-spec-2025} while isolating the effect of template edits on multi-turn cost amplification.

\end{document}